\documentclass[conference]{IEEEtran}
\IEEEoverridecommandlockouts
\pdfoutput=1
\usepackage{subfigure}
\usepackage{float}
\usepackage{stfloats}
\usepackage{multirow}
\usepackage{amsmath,amssymb,amsfonts}
\usepackage[ruled]{algorithm2e}
\usepackage{graphicx}
\usepackage{textcomp}
\usepackage{xcolor}
\usepackage[style=numeric, sorting=none]{biblatex}
\begin{document}
\title{Enhanced Preamble Based MAC Mechanism for IIoT-oriented PLC Network\\
% {\footnotesize \textsuperscript{*}Note: Sub-titles are not captured in Xplore and
% should not be used}
% \thanks{Identify applicable funding agency here. If none, delete this.}
}

\author{
 \IEEEauthorblockN{Kai Song, Biqian Feng, Yongpeng Wu and Wenjun Zhang}

 \thanks{K. Song, B. Feng, Y. Wu, W. Zhang are with the Department of Electronic Engineering, Shanghai Jiao Tong University, Minhang 200240, China (e-mail: gansusongkai@sjtu.edu.cn; fengbiqian@sjtu.edu.cn; yongpeng.wu@sjtu.edu.cn; zhangwenjun@sjtu.edu.cn) (Corresponding author: Yongpeng Wu).}
 
%  \thanks{X.-G. Xia is with the Department of Electrical and Computer Engineering, University of Delaware, Newark, DE 19716, USA. (e-mail: xxia@ee.udel.edu).}
 
%  \thanks{C. Xiao is with the Department of Electrical and Computer Engineering, Lehigh University, Bethlehem, PA 18015 USA (e-mail: xiaoc@lehigh.edu).}
}

\maketitle

\begin{abstract}
In this paper, we propose an enhanced preamble based media access control mechanism (E-PMAC), which can be applied in power line communication (PLC) network for Industrial Internet of Things (IIoT). We introduce detailed technologies used in E-PMAC, including delay calibration mechanism, preamble design, and slot allocation algorithm. With these technologies, E-PMAC is more robust than existing preamble based MAC mechanism (P-MAC). Besides, we analyze the disadvantage of P-MAC in multi-layer networking and design the networking process of E-PMAC to accelerate networking process. We analyze the complexity of networking process in P-MAC and E-PMAC and prove that E-PMAC has lower complexity than P-MAC. Finally, we simulate the single-layer networking and multi-layer networking of E-PMAC, P-MAC, and existing PLC protocol, i.e. , IEEE1901.1. The simulation results indicate that E-PMAC spends much less time in networking than IEEE1901.1 and P-MAC. Finally, with our work, a PLC network based on E-PMAC mechanism can be realized.
% Based on our work, a PLC network based on E-PMAC can be realized and deployed to IIoT.
\end{abstract}


\section{Introduction}
Nowadays, communication technology plays a more and more important role in industrial production. Industrial Internet of Things (IIoT) is considered as the foundation of Industry 4.0 and Intelligent Manufacturing \cite{IIoT4_0}. It fills the gap between the devices and information technology (IT) systems, and eliminates the phenomenon of information silos \cite{Silos}\cite{Silos2}.

IIoT is a network which is constructed by industrial data collectors. In IIoT, a large number of low-cost nodes, including sensors and meters, are deployed in the factories. These nodes serve as the underlayer of IIoT and collect various data.

The implementation of IIoT needs advanced communication technology. In the IIoT, the number of nodes is very large and the density of nodes is high, which is different from other networks. In this case, the cost and the time spending in network construction cannot be ignored. Therefore, the physical (PHY) layer and multi-access control (MAC) layer should be specifically designed.

\subsection{Related work}
There has been a lot of research on the PHY layer and MAC layer of IIoT. In terms of PHY layer, promising communication technologies include narrow band Internet of Things (NB-IoT), power line communication (PLC), etc. PLC has a unique advantage that it utilizes ubiquitous power line to communicate, which makes it more reliable than wireless communication in the factory scenario where blind areas of wireless signal are common. Although there exists attenuation and multipath propagation in power line \cite{multipath}, technologies like multiple-input multiple-output (MIMO) and orthogonal frequency division multiplexing (OFDM) are used in PLC \cite{MIMO_OFDM}, which improves communication quality. In \cite{IEEE1901}, IEEE1901.1 protocol is proposed and it can realize broadband PLC communication in grid system. In \cite{FDPLC}, frequency division power line communication (FD-PLC) is designed as a PHY layer protocol for IIoT-oriented PLC network to solve the problem that passband of PLC channel changes with time. All of these technologies make PLC a promising technology in IIoT.

As for MAC layer, because of the large number and high density of IIoT nodes, MAC layer protocol mainly aims at speeding up the networking process and carrier sense multiple access (CSMA) is widely used in MAC layer to decrease collision. The modelling of CSMA and network topology are main research directions \cite{Markov}\cite{cluster}. Besides of these research, an innovative preamble based MAC (P-MAC) mechanism is proposed in \cite{PMAC}. In P-MAC, channel competition and data transportation are totally divided and preamble is utilized to compete for access to central node. Compared with data frame, preamble is much shorter and can save networking time a lot.

To date, PLC protocol applied for the IIoT in factory remains to be designed. The power line channel in factories is noisier and passband frequency changes with time, which makes IEEE1901.1 infeasible. Most researches like \cite{Markov} and \cite{cluster} pay less attention to the speed of networking. The authors of \cite{PMAC} discuss networking speed and convey an idea of using preambles to establish connection. However, P-MAC in is designed for situation with a small amount of nodes. In aspect of realization, P-MAC needs receiver with high process speed, which raises the cost of devices and increases the probability of failure connection. In aspects of networking speed, nodes using P-MAC send more data frames than IEEE1901.1 in networking, which indicates P-MAC may have even lower speed than existing mechanisms. Therefore, a preamble based MAC mechanism with higher robustness and faster networking speed is desired for research.

\subsection{Our work}

Based on the idea of \cite{PMAC}, we propose enhanced P-MAC (E-PMAC), a MAC mechanism applied in IIoT-oriented PLC network. Compared with P-MAC, E-PMAC is more robust and have faster networking speed. Firstly, we introduce detailed technologies used in E-PMAC, including delay calibration mechanism, preamble designing, and slot allocation algorithm. With these technologies, E-PMAC is more robust than P-MAC in real IIoT environment. Secondly, we design the single-layer networking and multi-layer networking process of E-PMAC. Thirdly, we compare P-MAC and E-PMAC in the complexity of networking and prove that E-PMAC can spend less time in networking than P-MAC. Finally, we make simulations to compare E-PMAC, P-MAC, and IEEE1901.1 in networking time. The simulation results indicate that E-PMAC is faster than IEEE1901.1 and P-MAC.

The rest of this paper is organized as follows. Section II
introduces detailed technologies used in E-PMAC. In Sections III and IV, we introduce the networking process of E-PMAC and compare P-MAC and E-PMAC in complexity. Section V presents simulations of P-MAC, E-PMAC, and IEEE1901.1. Finally, we conclude the paper in Section VI.

\section{Technical Details in E-PMAC}
According to \cite{IEEE1901}, in IIoT based on PLC, the nodes can be divided into three types, including Central Coordinator (CCO), Proxy Coordinator (PCO), and Station (STA). The P-MAC mechanism includes three parts, preamble time exchange (PTE), T-Query handshake (T-Query), and network configuration (Net-Config) \cite{PMAC}.

In this section, we introduce delay calibration mechanism, preamble designing, and slot allocation algorithm, which make E-PMAC more robust than P-MAC.

\subsection{Delay Calibration Mechanism}
The basic idea of using preamble to establish connection is that CCO can match STA with time difference of two preambles. In PTE, CCO sends a preamble to STA and the STA responses CCO with another preamble. The time difference of these two preambles can be measured by both CCO and STA \cite{PMAC}. In order to complete the handshake in T-Query, it should be guaranteed that the time differences measured by CCO and STA is equal.

\begin{figure}[htbp]
\flushleft
\includegraphics[width=8.5cm]{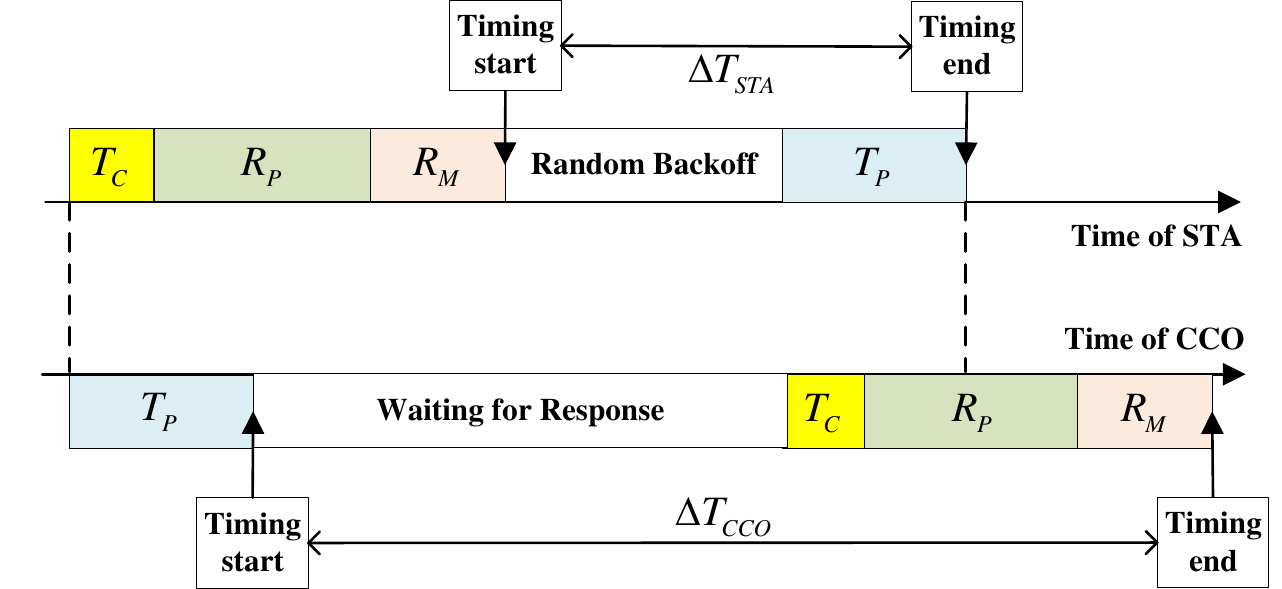}
\caption{Delay in PTE process}
\label{fig:deltaT}
\end{figure}

Sending delay, propagation delay, and receiving delay inevitably occur in practical PLC system. Besides, the data transportation time between PHY hardware and MAC hardware cannot be ignored. All kinds of delays are listed in TABLE \ref{tab:delay} and the process of time difference calculation is shown in Fig. \ref{fig:deltaT}, in which delays make $\Delta{T}_{STA}$ and $\Delta{T}_{CCO}$, time differences measured by STA and CCO different. In IIoT, the distance between nodes is usually small and the error caused by propagation delay is much less than that caused by other delays, so $T_{C}$ can be ignored, but other delays in TABLE \ref{tab:delay} are supposed to be considered. As for P-MAC, however, it only considers the situation that $T_{P}$ equals to $R_{P}$ plus $R_{M}$, which means that the performance of receiver should be good enough to have a processing delay less than the timing clock. This condition is difficult to meet because nodes in IIoT should be low-cost.

\begin{table}[htbp]
\caption{Hardware Delay in P-MAC}
\begin{center}
\begin{tabular}{|c|c|c|}
\hline
\textbf{Notation}&\textbf{Type of Delay}\\
\hline
$T_{C}$&Signal propagation in channel\\
\hline
$T_{P}$&Coding and transmitting data to channel\\
\hline
$R_{P}$&Receiving and decoding data from channel\\
\hline
$R_{M}$&Sending data from PHY to MAC\\
\hline
\end{tabular}
\label{tab:delay}
\end{center}
\end{table}

\begin{figure}[htbp]
% \flushleft
\centering
\includegraphics[width=6.5cm]{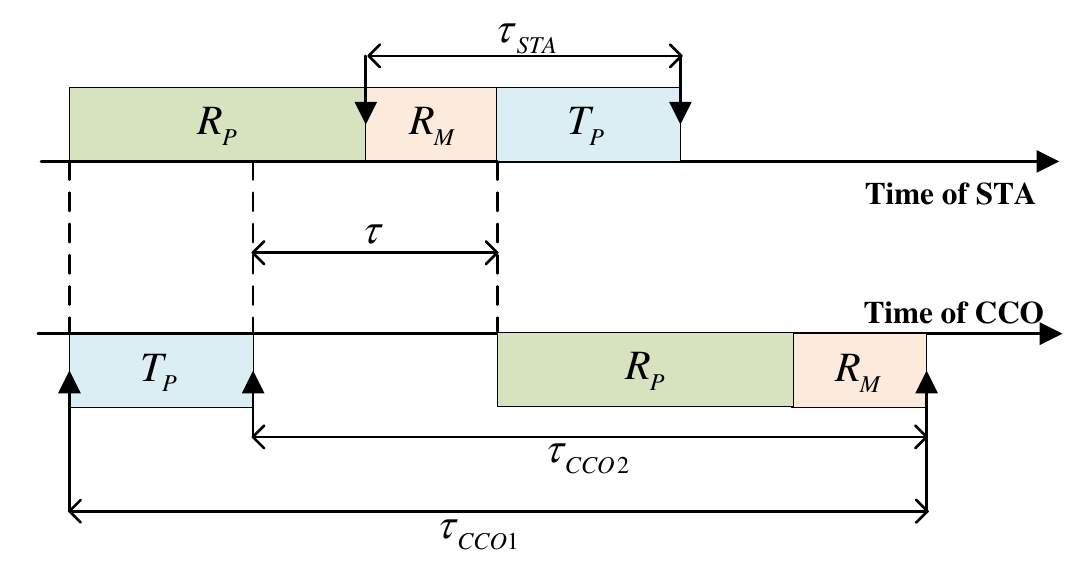}
\caption{Delay calibration mechanism of E-PMAC}
\label{fig:correction}
\end{figure}

We add delay calibration mechanism to E-PMAC so that CCO can still get correct time difference with non-real-time receiver, which makes E-PMAC more robust than P-MAC. In OFDM-based PLC systems, preamble and data frames have fixed length, so the $T_P$, $R_P$, and $R_M$ in Table-\ref{tab:delay} are constant. Based on this feature, we design the delay calibration mechanism shown in Fig. \ref{fig:correction}. The random backoff time of STA can be set to zero. CCO measures out the time differences $\tau_{CCO1}$ and $\tau_{CCO2}$. STA measures out $\tau_{STA}$, and sends it to CCO through subsequent data frame. By letting a correction factor $\tau$, linear equations about $T_{P}$, $R_{P}$, $R_{M}$ and $\tau$ is given by \eqref{eq:correction}. At the beginning of networking, CCO can solve \eqref{eq:correction} and save the value of $\tau$. Then in PTE, CCO can utilize \eqref{eq:corrected} to figure out $\Delta\hat{T}_{STA}$, which is equal to $\Delta{T}_{STA}$. In T-Query, CCO will send $\Delta\hat{T}_{STA}$ instead of $\Delta T_{CCO}$ to STA.

\begin{equation}
\left\{
\begin{aligned}
    % R_{P}+&R_{M}-T_{P}&-\Delta t &= 0\\
    % R_{P}+&R_{M}      &+\Delta t &= \Delta T_{CCO2} \\
    % R_{P}+&R_{M}+T_{P}&+\Delta t &= \Delta T_{CCO1}\\
    %       &R_{M}+T_{P}&          &= \Delta T_{STA}
    R_{P}+&R_{M}-T_{P}&-\tau &= 0\\
    R_{P}+&R_{M}      &+\tau &= \tau_{CCO2} \\
    R_{P}+&R_{M}+T_{P}&+\tau  &= \tau_{CCO1}\\
          &R_{M}+T_{P}&          &= \tau_{STA}
\end{aligned}
\right.
\label{eq:correction}
\end{equation}

\begin{equation}
    \Delta\hat{T}_{STA} = \Delta T_{CCO}-2\tau
\label{eq:corrected}
\end{equation}

\subsection{Preambles in E-PMAC}
P-MAC uses 2 kinds of preambles, while in E-PMAC, 4 kinds of preambles are used. The preamble used as head of data frame should be unique. In PTE, the downward preamble differs from the upward preamble. In Net-Config of P-MAC, STAs will send acknowledge (ACK) to CCO, so we need another kind of preamble used as ACK. We define the preambles in data frame, downward part of PTE, upward part of PTE and Net-Config as DAT, NET, REQ, and ACK respectively.

In PLC network, there exist some STAs which cannot receive signal from CCO but can communicate with other STAs. We call these STAs as hidden STAs. In the PTE of P-MAC, the upward and downward preambles are the same. If hidden STAs receive upward preambles sent from other STAs to CCO, they will treat these preambles as downward preambles and response upward preambles, which will cause more collision. While in E-PMAC, NET and REQ serve as downward and upward preamble, eliminating the collision caused by hidden STAs.

\subsection{PTE Slot Allocation Algorithm}
We define one PTE, T-Query, and Net-Config as one networking cycle (NC). In \cite{PMAC}, P-MAC is only used in the process of CCO networking with 2 STAs in one NC. In IIoT, the number of STAs is so large that collision in PTE is inevitable and CCO cannot connect with all the STAs in only one NC. The problems that E-PMAC faces are listed as follows.
\begin{itemize}
    \item Collision rate of preamble is high because CSMA cannot be used in PTE. In PTE process, only the preamble is sent, and STA cannot detect preamble until the transmission of preamble is finished. Therefore, STAs cannot detect that the channel is busy, so they can only send REQ preamble to CCO after waiting for a random time without collision detection. There will be a high probability of collisions in the process.
    \item The exact number of STAs is unknown for CCO, and CCO can only allocate a random number of slots. If the number of slots is too small, collision will be too frequent for STA to send preamble to CCO. And if the number of idle slots is too large, networking will be too slow.
    % \item The number of time slots in PTE should dynamically change. At the beginning of networking, the exact number of STAs is unknown for CCO, and CCO can only allocate a random number of slots. If the number of slots is too small, collision will be too frequent for STA to send preamble to CCO. And if the number of idle slots is too large, networking will be too slow. Hence, the number of slots should be changeable and close to the number of STAs.
\end{itemize}

To solve the two problems, we design a PTE slot allocation algorithm for E-PMAC. The basic idea is to adjust the allocated slots by referring to the recorded number of slots and STAs successfully joining network in the last NC. The algorithm is shown in \emph{Algorithm \ref{alg:slot}}.

In the PTE of the first NC, we replace NET preamble with a data frame containing the number of slots, so that the STAs can know the slots number in the first PTE. In the subsequent NCs, CCO and STAs figure out number of slots by \emph{Algorithm \ref{alg:slot}}. \emph{Algorithm \ref{alg:slot}} requires that STAs should know the number of STAs successfully joining network in the last NC. We introduce the solution to this problem in section \ref{TDF}. By \emph{Algorithm \ref{alg:slot}}, E-PMAC can dynamically increase slots number when collision rate is too high and keep the slots number close to STAs number. Because the data used in \emph{Algorithm \ref{alg:slot}} have been stored, the algorithm just needs to run a judgement process and has a complexity of O(1).

\begin{algorithm}
% \SetAlgoNoLine  %去掉之前的竖线
\caption{Allocation Algorithm of PTE Slot}
\label{alg:slot}
\KwIn{\\
$N_{slot}$: Slots number in the last PTE\\
$N_{STA}$: STAs number in the last PTE\\
$T_{f}$: Times of PTE that no STAs joining\\
$T_{PTE}$: Times of finished PTE\\
}
\KwOut{\\
$N_{slot}^{\prime}$: Slots number in the next PTE
}
\BlankLine
Initialize $N_0,T_{fMax}\in\mathbb{N}^+$, $\eta_{min}\in(0, 1)$, $K_1,K_2>1(K_1<K_2)$;
        
\eIf{$T_{PTE}\neq0$}
{
    $\eta=N_{STA}/N_{slot}$;\\
    \eIf{$\eta>0$}
    {
        \eIf{$\eta\leq\eta_{min}$}
        {
            $N_{slot}^{\prime}=K_1\times N_{slot}$;
        }
        {
            $N_{slot}^{\prime}=N_{slot}$;
        }
    }
    {
        \eIf{$T_{f}\leq T_{fMax}$}
        {
            $N_{slot}^{\prime}=K_2\times N_{slot}$;
        }
        {
            $N_{slot}^{\prime}=0$;
        }
    }
}
{
    $N_{slot}^{\prime}=N_0$;
}
return $N_{slot}^{\prime}$;
\end{algorithm}

\section{Networking Process of E-PMAC}
When comparing P-MAC with IEEE1901.1, we can find in the networking process of IEEE1901.1, once a STA successfully sends data frame containing information necessary for networking to CCO by CSMA, CCO will let the STA join the network without requiring it for other information. Although P-MAC uses preamble to save more time in competing access to CCO than IEEE1901.1, it needs more data frames to send information. Hence, the networking time of P-MAC still has probability to be longer than IEEE1901.1.

In the scheme of P-MAC, CCO and STAs only send one datum in one data frame, which wastes lots of space. From this perspective, we design the networking process of E-PMAC based on the idea of making full use of data frames.

\subsection{Single-layer Networking of E-PMAC}\label{TDF}
For T-Query of E-PMAC, we define a new frame called Time Differences Frame (TDF), in which time differences of STAs are sorted in ascending order. CCO sends TDFs at the beginning of T-Query. The length of TDF is constant and CCO will send multiple TDFs if there are too many time differences. When STAs receive TDFs, they can get the ascending order of time differences, and then send MAC addresses to CCO in the same order.

In Net-Config of E-PMAC, pairs of short ID (SID) and MAC address are sent from CCO to STAs. So we define a new frame called SIDs Frame (SDF), which is filled with pairs of short ID (SID) and MAC address of STA. In Net-Config, CCO sends SDFs and STAs can get whether they join in the network successfully. The successful STAs send ACK to CCO at the end of Net-Config. Besides, by SDF, STAs can know the number of successful STAs and calculate the PTE slots number in the next NC with \emph{Algorithm \ref{alg:slot}}.

\begin{figure}[htbp]
\centering
\includegraphics[width=8.0cm]{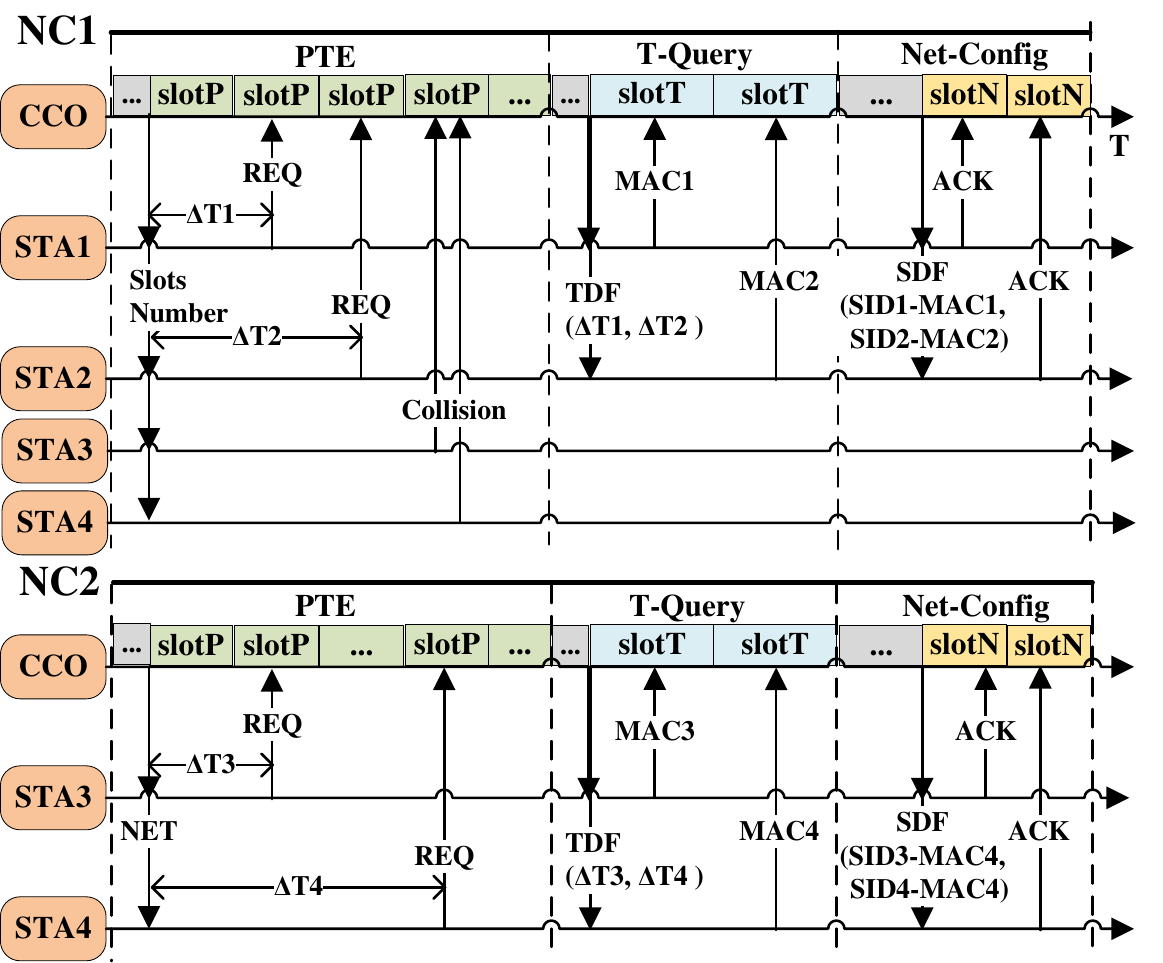}
\caption{Single-layer networking of E-PMAC}
\label{fig:IPMAC-single}
\end{figure}

Considering a scenario that CCO connects with 4 STAs in 2 NCs, the process of E-PMAC is shown in Fig. \ref{fig:IPMAC-single}. In PTE of the NC1, CCO sends data frame containing slots number. STA1 and STA2 send REQ in different slots and join in network successfully. STA3 and STA4 send REQ in the same slot and cause collision. When collision happens, although the two preambles partially overlap, CCO can still distinguish them by correlation detection and detect the collision. CCO will not send TDF containing time difference of STA3 and STA4 in T-Query. In Net-Config, all the STAs get the number of successful STAs by SDF and they use \emph{Algorithm \ref{alg:slot}} to calculate the slots number of PTE in NC2. In NC2, CCO just send NET preamble in PTE, and STA3 and STA4 join in the network in the same way as STA1 and STA2.

\subsection{Multi-layer Networking of E-PMAC}
The Beacon is necessary in multi-layer networking. In \cite{PMAC}, beacon is a data frame that contains instruction of PCO starting networking. It is obvious that CCO knows the largest SID at the moment. Hence, in E-PMAC, CCO adds the largest SID to beacon and send beacon to PCO. In this way, PCO gets the largest SID and it can allocate SID to STAs independently, instead of requesting SIDs from CCO. After the PCO connecting with STAs, it just sends the SIDs and MAC addresses of STAs to CCO and CCO updates the network information.

\begin{figure}[htbp]
\centering
\includegraphics[width=8.0cm]{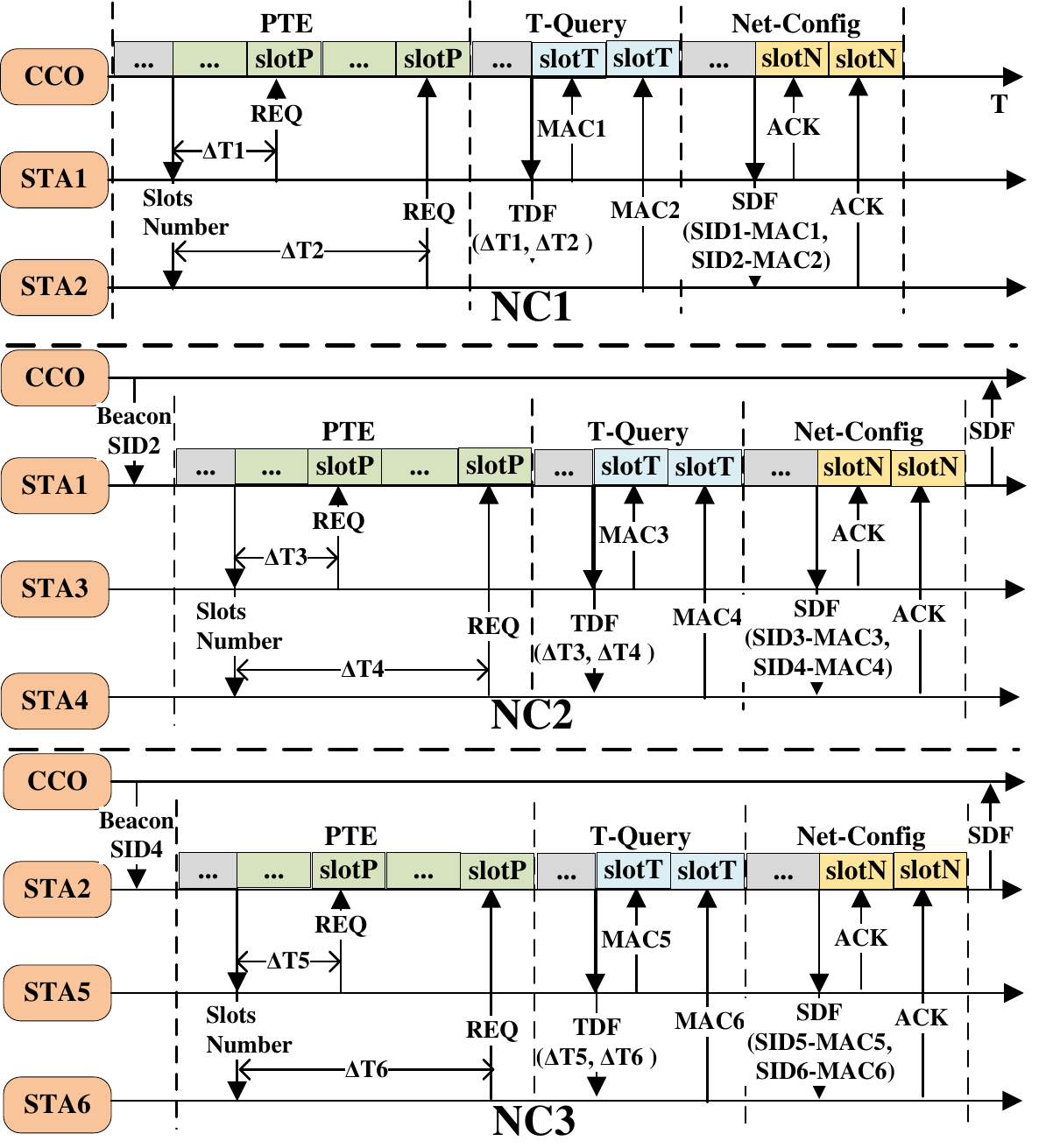}
\caption{Multi-layer networking of E-PMAC}
\label{fig:IPMAC-multi}
\end{figure}

Considering a scenario that CCO connects with 6 STAs in 2 layers, the process of E-PMAC is shown in Fig. \ref{fig:IPMAC-multi}. CCO can only communicate with STA1 and STA2. In NC1, STA1 and STA2 connect with CCO. In NC2, CCO sends beacon containing SID2, the currently largest SID, to STA1. STA1 serves as PCO, connects with STA3 and STA4, and send SDF to CCO. CCO can update the information of STA3 and STA4. In NC3, CCO makes STA2 serve as PCO and connect with STA5 and STA6.

\section{Complexity Analysis of P-MAC and E-PMAC}
Networking mechanism should be designed to send data frames as few as possible in the process of networking. In this section, the times of sending data frames is adopted as a metric to evaluate the complexity of networking for P-MAC and E-PMAC. Our analysis shows that E-PMAC has lower complexity than P-MAC and this advantage will be more obvious in multi-layer networking.

\subsection{Networking Complexity of P-MAC}\label{sub:Pmac}
\subsubsection{Single-layer Networking}
According to the design of \cite{PMAC}, in T-Query process and Net-Config process, time difference measured by CCO, MAC address of STA and SID are sent between STA and CCO in three data frames respectively. Let $N$ be the number of STAs and the number of data frames is $3N$.

\subsubsection{Multi-layer Networking}
In practice, the CCO may not be able to communicate with all the STAs and the network may have multiple layers. In this case, CCO sends beacon to PCO and controls PCO to start PTE. P-MAC requires that PCO should send beacon, which contains time difference of a STA, to CCO after PTE. In T-query and Net-Config, CCO and the STA exchange time difference, MAC address, and SID through the PCOs which compose the path from CCO to STA.

Therefore, let $K$ be the number of layers and $M$ be the number of STAs with which a PCO can communicate. Then the network structure becomes a $M$-ary tree with a depth of $K$. Let $n_{k}$ be the number of data frames sent in the process of CCO controlling one PCO in ($k-1$)-th layer to connect with STAs in the $k$-th layer by P-MAC. The recurrence formula of $n_{k}$ is given by \eqref{eq:pmac-multi1}. Then, $N(M,K)$, the number of data frames sent in the multi-layer networking of P-MAC, is given by \eqref{eq:pmac-multi2}.

\begin{equation}
\begin{aligned}
    n_{k}&=
    \left\{
    \begin{aligned}
        &n_{k-1}+3M+2,k\in\{2,\cdots,K\}\\
        &3M,k=1
    \end{aligned}
    \right.\\
\end{aligned}
\label{eq:pmac-multi1}
\end{equation}

\begin{equation}
\begin{aligned}
    N(M,K)&=\sum_{k=1}^{K}n_{k}\times{M^{k-1}}\\
    % &=\sum_{k=1}^{K}[(3M+2)k-2]M^{k-1}\\
    &=(AK+B)M^K-B\\
A&=3+5/(M-1),B=-5M/(M-1)^2
\end{aligned}
\label{eq:pmac-multi2}
\end{equation}

\subsection{Networking Complexity of E-PMAC}
\subsubsection{Single-layer Networking}
In practice, OFDM data frame payload ranges from 100 bytes to 200 bytes. The length of time difference is less than 5 bytes and the length of the pair of SID and MAC address is less than 10 bytes. Therefore, SDF contains over 10 SID-MAC pairs and TDF contains over 20 time differences. For the networking of $N$ STAs, the number of data frames sent in T-Query and Net-Config is $\lceil0.1N\rceil+\lceil0.05N\rceil+N$, where ``$\lceil\cdot\rceil$" is ceiling operation.

\subsubsection{Multi-layer Networking}
We simplify the number data frames in single-layer networking to $N+2$. Considering the same multi-layer network as \ref{sub:Pmac}, $n^{\prime}_{k}$, the number of data frames in E-PMAC sent in the process of CCO controlling one PCO in ($k-1$)-th layer connecting with STAs in the $k$-th layer by E-PMAC, is given by \eqref{eq:Ipmac-multi1}. And $N^{\prime}(M,K)$, the number of data frames in the whole multi-layer networking of E-PMAC, is given by \eqref{eq:Ipmac-multi2}. 

With $N(M,K)$ and $N^{\prime}(M,K)$, we can get $\Delta_{STA}$, the reduction of data frames in the process of CCO connecting with one STA, given in \eqref{eq:Ipmac-multi3}. Therefore, E-PMAC has a lower complexity and E-PMAC can save much more time than P-MAC in multi-layer networking process with a large number of STAs.

\begin{equation}
\begin{aligned}
    n^{\prime}_{k}&=
    \left\{
    \begin{aligned}
        &n^{\prime}_{k-1}+2,k\in\{2,\cdots,K\}\\
        &M+2,k=1
    \end{aligned}
    \right.\\
\end{aligned}
\label{eq:Ipmac-multi1}
\end{equation}

\begin{equation}
\begin{aligned}
    N^{\prime}(M,K)&=\sum_{k=1}^{K}n^{\prime}_{k}\times{M^{k-1}}
    =\sum_{k=1}^{K}(2k+M)M^{k-1}\\
    &=(A^{\prime}K+B^{\prime})M^K-B^{\prime}\\
    A^{\prime}&=2/(M-1),
    B^{\prime}=1+(M-3)/(M-1)^2
\end{aligned}
\label{eq:Ipmac-multi2}
\end{equation}

\begin{equation}
\begin{aligned}
    \Delta_{STA}&=(N(M,K)-N^{\prime}(M,K))/[\frac{M(M^K-1)}{M-1}]\\
    &\approx(N(M,K)-N^{\prime}(M,K))(M-1)/M^{K+1}\\
    &\approx[(A-A^{\prime})K+B-B^{\prime}](M-1)/M\\
    &=3K-1-(5M-2)/(M^2-M)\approx3K-1
\end{aligned}
\label{eq:Ipmac-multi3}
\end{equation}

\section{Simulation}
We simulate the networking time of P-MAC, E-PMAC, and IEEE1901.1 in the scenario that a large number of STAs form a single-layer or a multi-layer network. According to the simulation results, E-PMAC can greatly speed up the process of networking compared with IEEE1901.1 and P-MAC.

\subsection{Parameters of PHY Layer}\label{parameters}
The use of P-MAC and E-PMAC needs PHY layer protocal. According to \cite{FDPLC}, FD-PLC is a PHY protocol which can configure central frequency and overcome the time-varying passband frequency in power line of factory. Hence, We combine FD-PLC and PMAC/E-PMAC as a complete protocol and make comparison with IEEE1901.1.

\subsubsection{Parameters of IEEE1901.1}
For IEEE1901.1, central beacon, proxy beacon, association request message (MMeAssocReq) and association indication message (MMeAssocInd) are mainly involved in networking \cite{IEEE1901}. Central beacon and proxy beacon are 127 bytes while MMeAssocReq and MMeAssocInd are 142 bytes. The beacons and messages are filled to 136 bytes and 272 bytes respectively and sent to PHY layer. Turbo coding, scrambling, channel interleaving, and ROBO (Robust OFDM) interleaving expand the bit length of beacons and messages to 21760 and 43520. The time length $T$ can be calculated by \eqref{eq:ieee1901}, in which $N_{s}$ is the number of OFDM symbol and $N_{c}$ is the number of feasible subcarriers. In actual industrial condition, available bandwidth is about 3MHz and $N_{c}$ is 94. Finally, the time length of beacons and messages are 9102us and 17488us, respectively.

\begin{equation}
\begin{aligned}
    T=&40.96\times(13+N_{s})+18.32\times2\\
    &+(N_{s}-2)\times10.8(us),N_{s}=\lceil{N_{b}}/{N_{c}}\rceil
\end{aligned}
\label{eq:ieee1901}
\end{equation}

\subsubsection{Parameters of FD-PLC}
In FD-PLC, data frames have 100 bytes. The PHY layer of FD-PLC includes convolutional coding, RS coding and ROBO interleaving \cite{FDPLC}, which expand data frame to 8768 bits. FD-PLC uses 1024-point OFDM symbols for data frames and 64-point OFDM symbols for preamble. The bandwidth and sampling frequency are 1.25MHz and 40MHz, then the time length of these two OFDM symbols become 51.2us and 819.2us respectively. For P-MAC and E-PMAC, the preamble consists of 5 OFDM symbols and costs 256us. The available subcarriers number is 720, so the time of data frames with 8768 bits is $256+8768/720\times819.2=10232us$.

Accordingly, the preamble and data frame of IEEE1901.1 and FD-PLC have similar time length. The simulation result of comparing IEEE1901.1 with P-MAC and E-PMAC based on FD-PLC can verify the advantage of E-PMAC to P-MAC and IEEE1901.1.

\subsection{Design of Simulation}
\subsubsection{Slot time of frame and preamble}
Based on the time length in \ref{parameters}, we set the time of slots as Table \ref{tab:parameters}. For fairness, the slot time of MMeAssocReq, MMeAssocInd, and Data Frame are set to be the same.

\begin{table}[htbp]
\caption{Slot Time of Frames and Preamble}
\begin{center}
\begin{tabular}{|c|c|c|c|}
\hline
\textbf{PHY Layer}& \textbf{Frame Type}& \textbf{Time Length(us)}& \textbf{Slot Time (us)}\\
\hline
\multirow{4}{*}{IEEE1901.1}& Central Beacon& 9102& 12000\\
\cline{2-4}& Proxy Beacon & 9102 & 12000\\
\cline{2-4}& MMeAssocReq & 17488 & 20000\\
\cline{2-4}& MMeAssocInd & 17488 & 20000\\
\hline
\multirow{2}{*}{FD-PLC}& Preamble& 256& 400\\
\cline{2-4}& Data Frame & 11000 & 20000\\
\hline
\end{tabular}
\label{tab:parameters}
\end{center}
\end{table}

\subsubsection{Simulation algorithm}
Networking process consists of NCs. In every NC, some STAs join in the network. The entire simulation is to simulate one NC, accumulate the time of NC, and repeat these two steps until no STA remains out of network.
% The simulation algorithm of NC in IEEE1901.1 is shown as \emph{Algorithm \ref{alg:IEEE1901}}. 
The simulation of NC in IEEE1901.1 is more complicated than that in P-MAC or E-PMAC, because IEEE1901.1 uses P-persistent CSMA mechanism. The simulation algorithm of NC in P-MAC and E-PMAC is just to randomly assign STAs to slots. The situation that STAs are assigned to the same slot is regarded as collision. Then we can obtain the number of successful STAs.

We model the topology of multi-layer network to be tree network. To make the network structure flexible and have enough branches, we set the maximum number of layers and the minimum number of STAs in the first layer. The simulation process of multi-layer network is to generate a network structure randomly and simulate the networking process layer by layer.

\subsubsection{Parameters of Simulation}
The important parameters relevant to simulation are shown in TABLE \ref{tab:Parameters}, where $N$ is the number of STAs and $K$ is the number of layers.

\begin{table}[H]
\caption{Parameters of Simulation}
\begin{center}
\begin{tabular}{|c|c|}
\hline
\textbf{Parameter}&\textbf{Value}\\
\hline
$K_1$ in \emph{Algorithm \ref{alg:slot}}&1.3\\
\hline
$K_2$ in \emph{Algorithm \ref{alg:slot}}&2\\
\hline
Coefficient of P-persistent CSMA &0.75\\
\hline
Maximum layer of network& 6\\
\hline
Minimum STAs in the first layer&$\lceil e^{\frac{\ln{N}}{K}}\rceil$\\
\hline
\end{tabular}
\label{tab:Parameters}
\end{center}
\end{table}

\subsection{Simulation of Single-layer Networking}
In the networking process, $N_{node}$, the number of STAs influences the networking time. Besides, $N_{slot}/N_{node}$, the ratio of the number of STAs and slots changes the rate of collision and influences networking time. Therefore, we set the $N_{node}$ in one layer to range from 50 to 650 and $N_{slot}/N_{node}$ to range from 0.5 to 2 and simulate the networking of P-MAC, E-PMAC, and IEEE1901.1 under different $N_{node}$ and $N_{slot}/N_{node}$.

\begin{figure*} [t!]
	\centering
	\subfigure[\label{fig:S_sub_ieee}IEEE1901.1]{
		\includegraphics[scale=0.38]{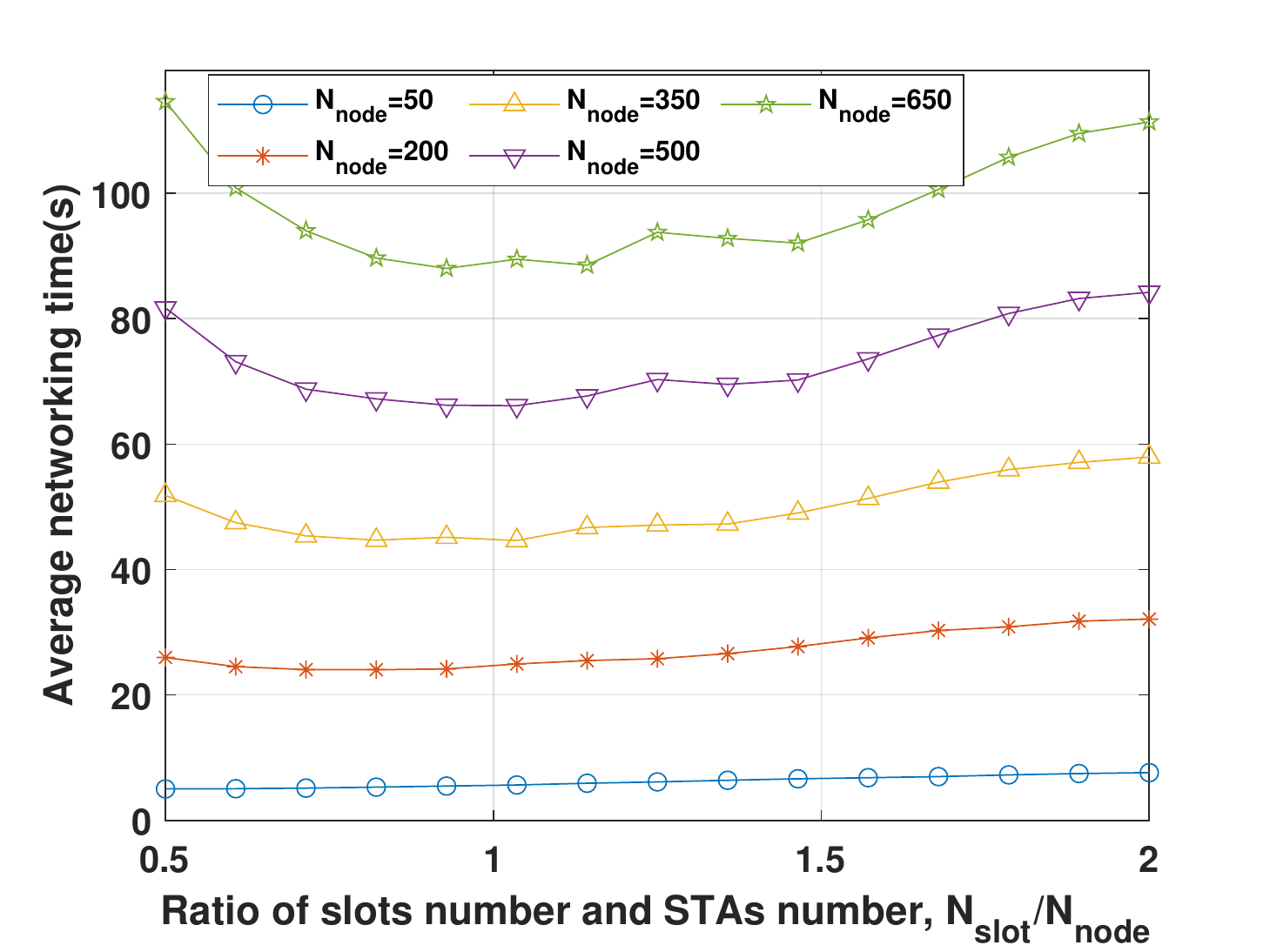}}
	\subfigure[\label{fig:S_sub_pmac}P-MAC]{
		\includegraphics[scale=0.38]{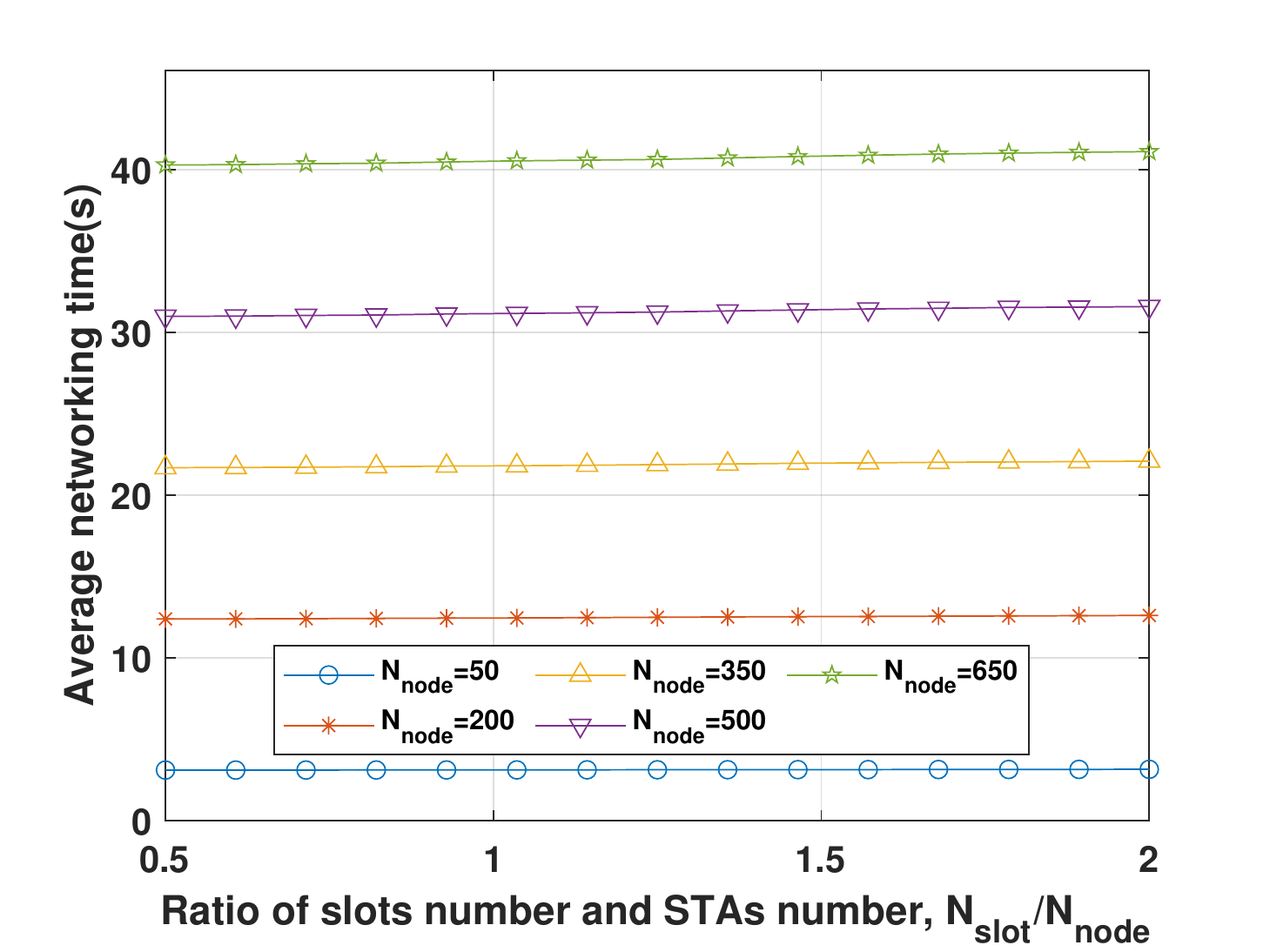}}
	\subfigure[\label{fig:S_sub_epmac}E-PMAC]{
		\includegraphics[scale=0.38]{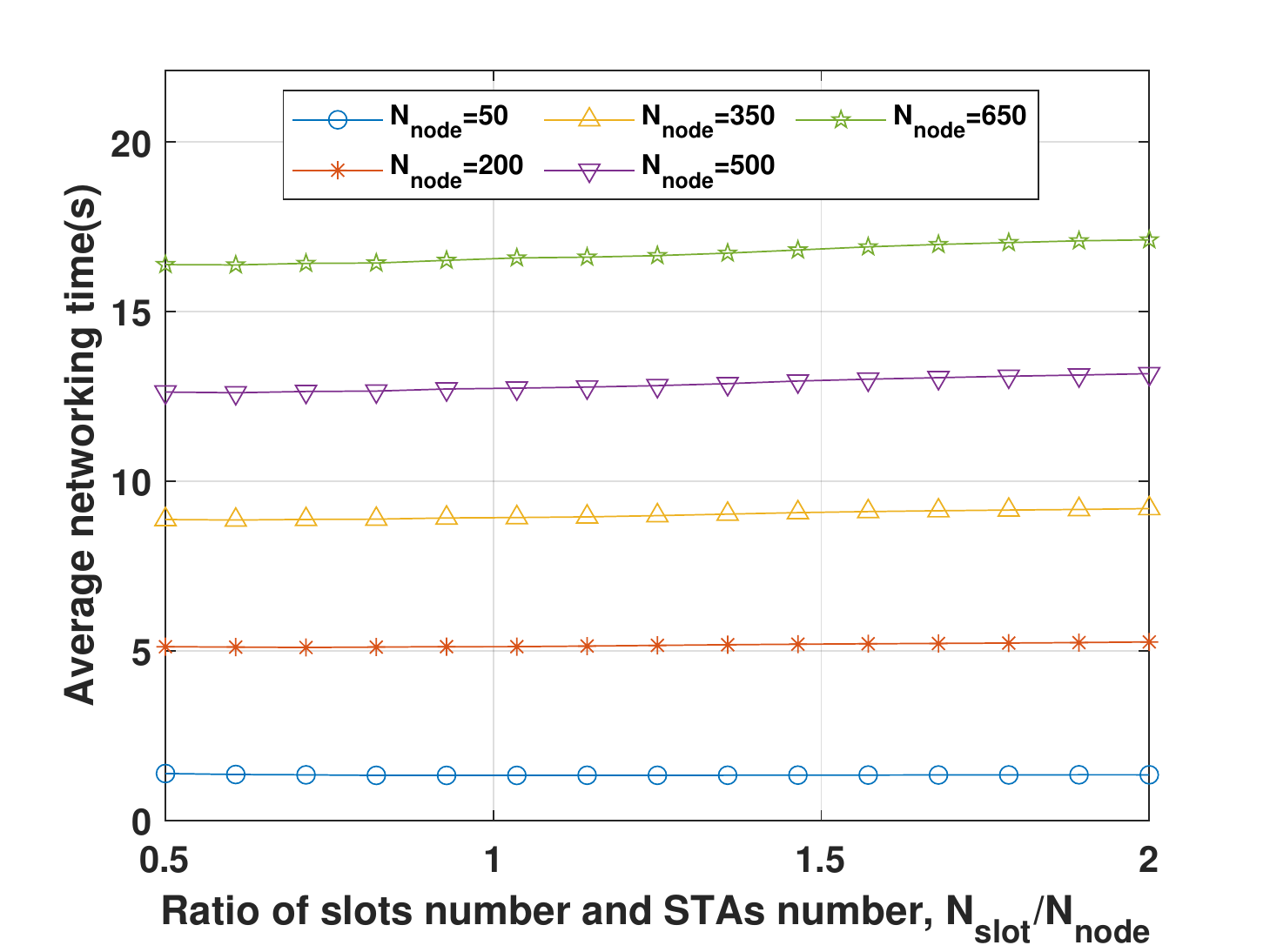}}
	\caption{Average networking time versus ratio of slots number and STAs number, $N_{slot}/N_{node}$}
	\label{fig:idle_rate} 
\end{figure*}

\begin{figure*}[h]
\centering
    \subfigure[\label{fig:single_min}Single-layer networking]{
		\includegraphics[scale=0.38]{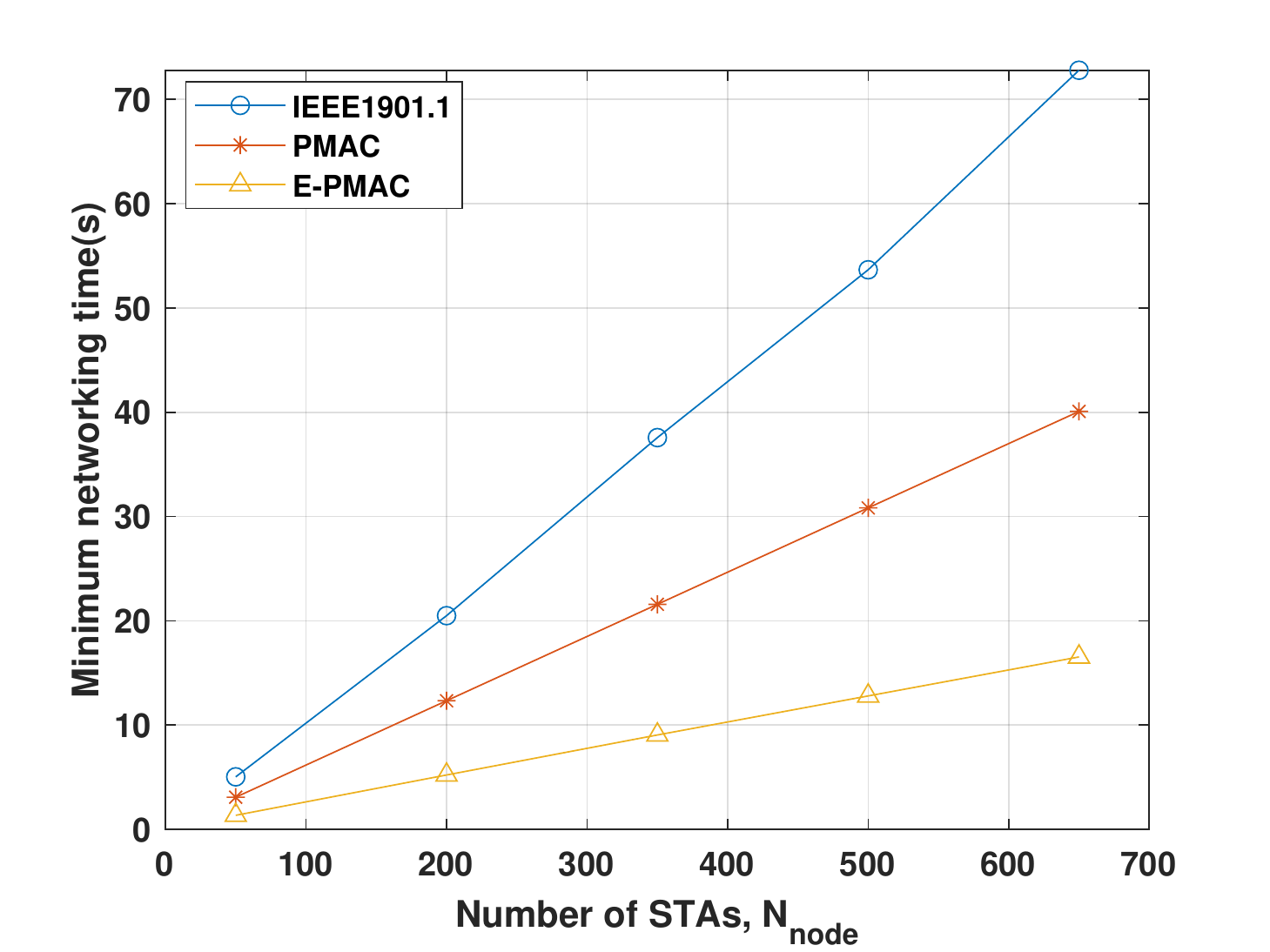}}
	\subfigure[\label{fig:multi_min}Multi-layer networking]{
		\includegraphics[scale=0.38]{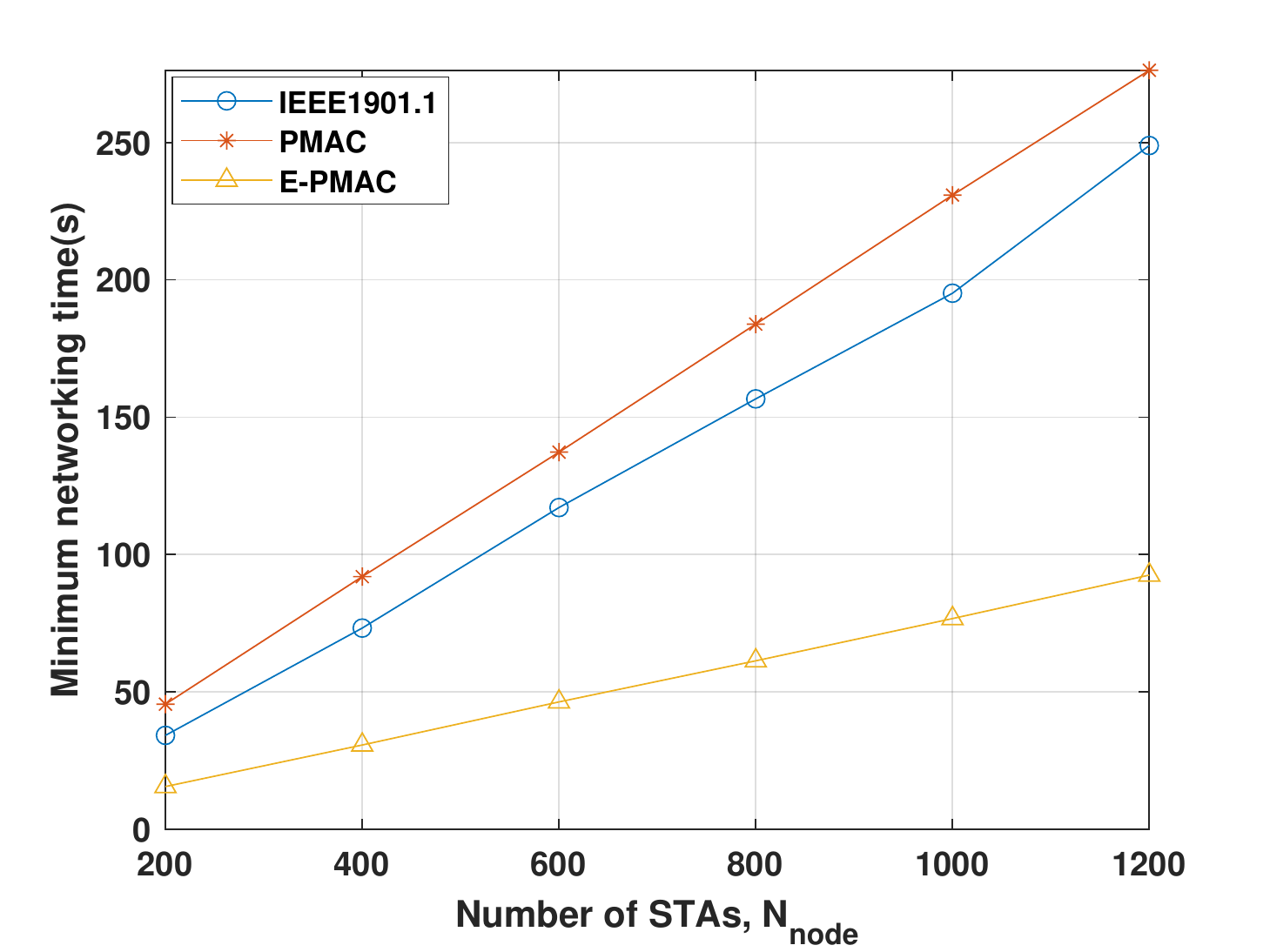}}
\caption{Minimum networking time versus number of STAs, $N_{node}$}
\label{fig:min_0_75}
\end{figure*}

In Fig. \ref{fig:idle_rate}, we compare average networking time of the three mechanisms versus $N_{slot}/N_{node}$. We can see that the networking time of IEEE1901.1 is more sensitive to $N_{slot}/N_{node}$ than P-MAC or E-PMAC. The minimum averaging networking time can be get in Fig. \ref{fig:idle_rate}, and then we compare the minimum networking time versus $N_{node}$ in Fig. \ref{fig:single_min}. It shows that E-PMAC can save more time than P-MAC and IEEE1901.1 in single-layer networking.

\subsection{Simulation of Multi-layer Networking}
Multi-layer networking is closer to reality and in multi-layer networking, the number of STAs will be larger. So we set the number of STAs to range from 200 to 1200 and do the same simulation as single-layer networking.

Considering the best situation, we assume that the $N_{slot}/N_{node}$ for shortest average networking time can be calculated in advance. We compare minimum average networking time of P-MAC, E-PMAC, and IEEE1901.1 versus STAs number in Fig. \ref{fig:multi_min}. We can see that the networking time of P-MAC even becomes larger than IEEE1901.1, which verifies that the P-MAC in \cite{IEEE1901} sends too many data frames and loses the advantage to IEEE1901.1. Compared with P-MAC, E-PMAC can still save a lot of time in multi-layer networking.
% \begin{figure}[htbp]
% \centering
% \includegraphics[width=8.5cm]{multi_cr_75.pdf}
% \caption{Mimimun average time under different $N_{STA}$}
% \label{fig:multi_best_0_75}
% \end{figure}

\begin{figure*} [ht]
	\centering
	\subfigure[\label{fig:sub_ieee}IEEE1901.1]{
		\includegraphics[scale=0.38]{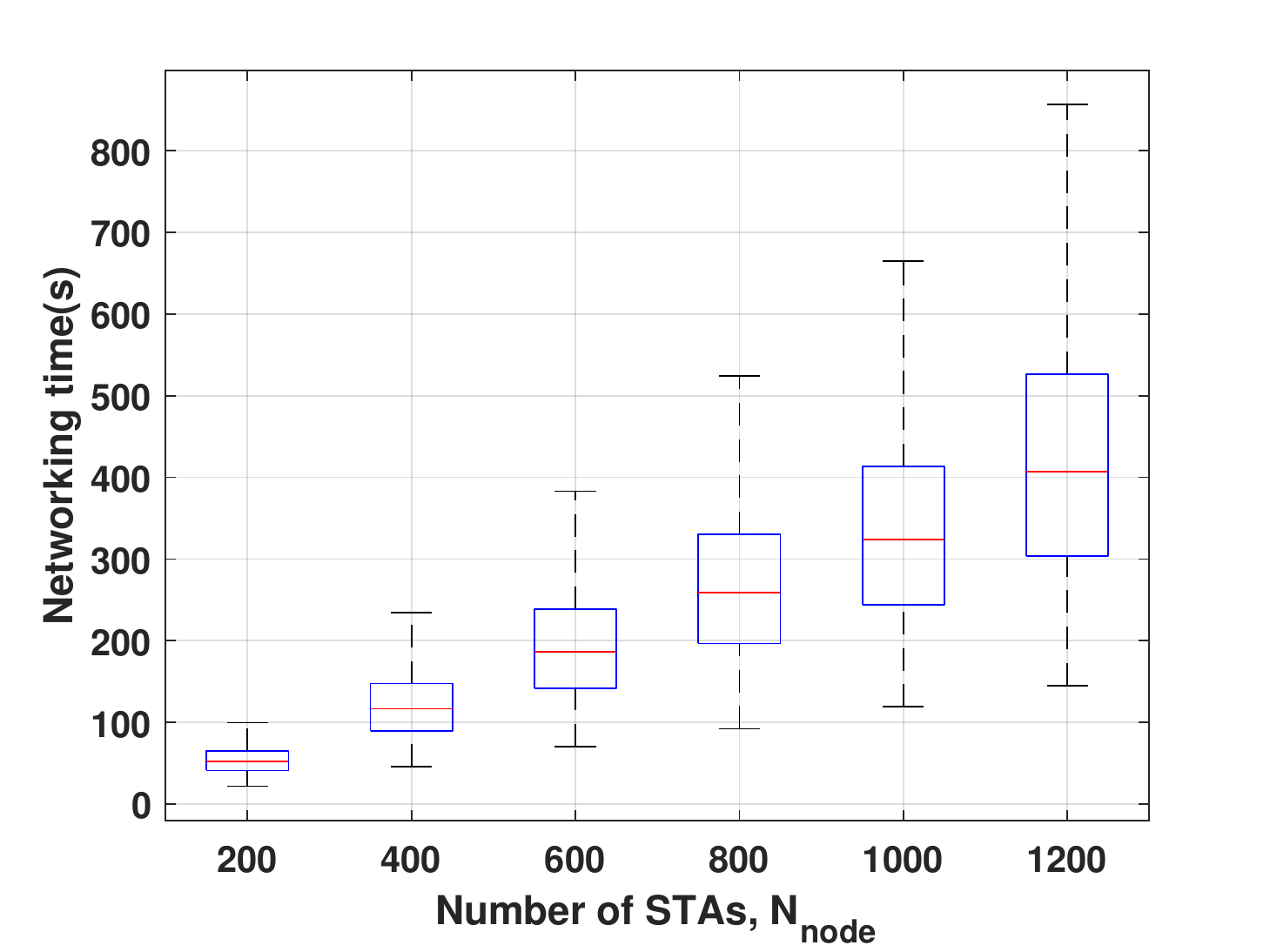}}
	\subfigure[\label{fig:sub_pmac}P-MAC]{
		\includegraphics[scale=0.38]{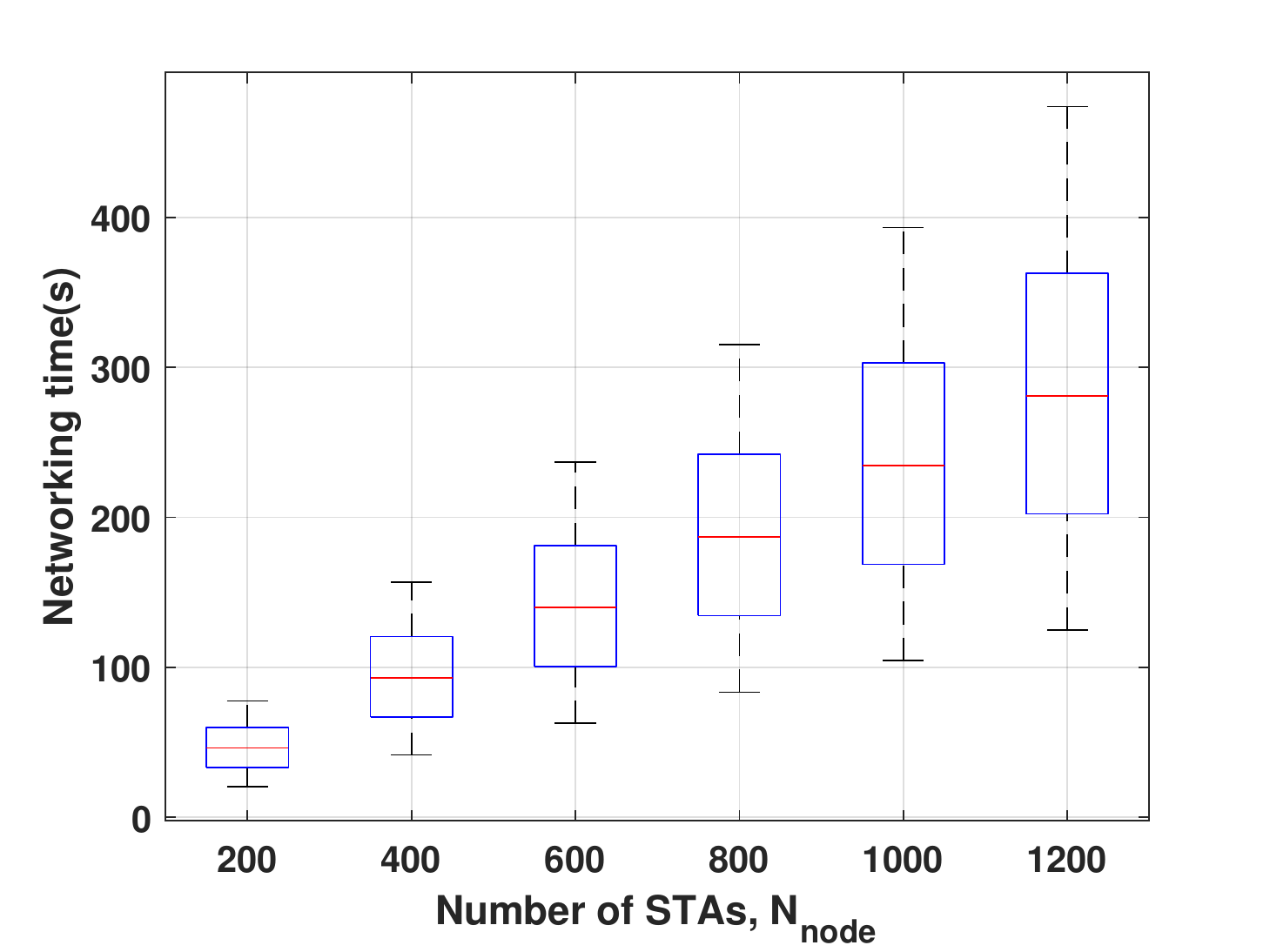}}
	\subfigure[\label{fig:sub_epmac}E-PMAC]{
		\includegraphics[scale=0.38]{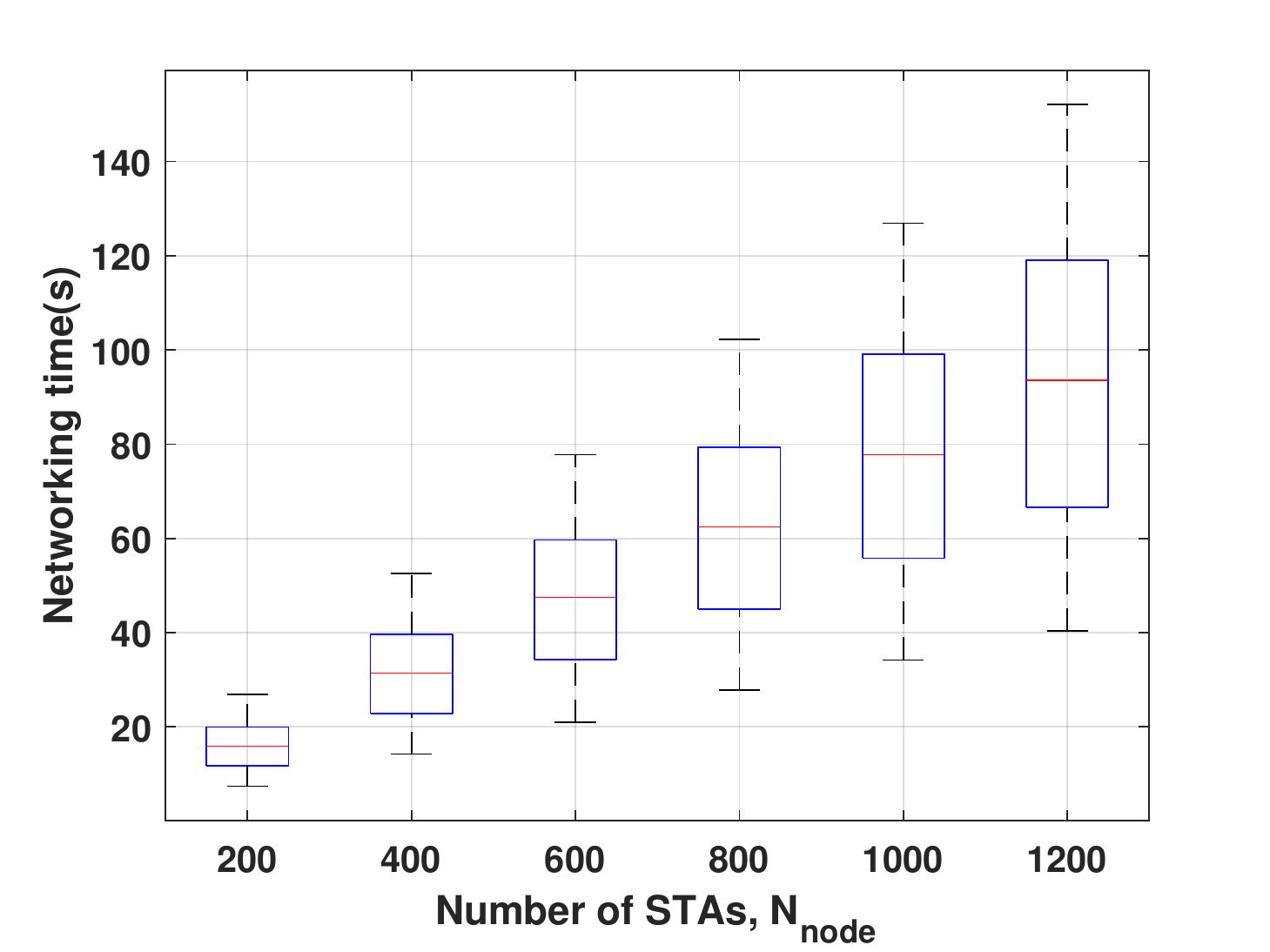}}
	\caption{Box plot of networking time versus STAs number, $N_{node}$}
	\label{fig:multi_box} 
\end{figure*}

In general situation, $N_{slot}/N_{node}$ changes randomly. We set $N_{slot}/N_{node}$ to randomly change between 0.5 and 2, and compare networking time of the three mechanisms versus $N_{node}$ in Fig. \ref{fig:multi_box}. The minimum networking time of P-MAC is approximately equal to that of IEEE1901.1, which is consistent to the result in Fig .\ref{fig:single_min}. Compared with P-MAC and IEEE1901.1, E-PMAC requires much less networking time. Besides, the difference of the longest and the shortest networking time is the smallest in E-PMAC, which indicates the networking time of E-PMAC is most steady when $N_{slot}/N_{node}$ changes randomly.

\section{Conclusion}
In this paper, we do further research on preamble based MAC mechanism and propose E-PMAC, an enhanced P-MAC used for IIoT-oriented PLC network. We add delay calibration mechanism, extra preamble types and slot allocation algorithm to make E-PMAC robust in IIoT. Besides, we find that P-MAC wastes lots of space in data frames and may not be faster in networking than existing protocols like IEEE1901.1. So we design the networking process of E-PMAC to overcome the disadvantages of P-MAC. We use the number of data frames sent in networking process as a metric of complexity and evaluate P-MAC and E-PMAC. Our analysis indicates that E-PMAC has lower complexity and can spend less time in networking theoretically. The simulation results verify that P-MAC will lose its advantage to IEEE1901.1 in multi-layer networking and E-PMAC spends much less time in networking than both P-MAC and IEEE1901.1.

\vspace{12pt}
% \color{red}
% IEEE conference templates contain guidance text for composing and formatting conference papers. Please ensure that all template text is removed from your conference paper prior to submission to the conference. Failure to remove the template text from your paper may result in your paper not being published.

\end{document}